\documentstyle[12pt]{article}
\def\bib{\bibitem}
\def\be{\begin{equation}}
\def\ee{\end{equation}}
\def\barr{\begin{array}}
\def\earr{\end{array}}

\def\GeV{\:{\rm GeV}}

\def\rp{$R_p \hspace{-1em}/\;\:$}

\def\ib#1,#2,#3{           {\it ibid.\/ }{\bf #1} (19#2) #3}
\def\ap#1,#2,#3{           {\it Ann. Phys. (NY)\/ }{\bf #1} (19#2) #3}
\def\ijmp#1,#2,#3{         {\it Int. J. Mod. Phys.\/ } {\bf A#1} (19#2) #3}
\def\mpl#1,#2,#3 {          {\it Mod. Phys. Lett.\/ } {\bf A#1} (19#2) #3}
\def\np#1,#2,#3{           {\it Nucl. Phys.\/ }{\bf B#1} (19#2) #3}
\def\npps#1,#2,#3{         {\it Nucl. Phys. B (Proc. Suppl.)\/ }{\bf B#1}
                             (19#2) #3}
\def\plb#1,#2,#3{           {\it Phys. Lett.\/ }{\bf B#1} (19#2) #3}
\def\pr#1,#2,#3{           {\it Phys. Rev.\/ }{\bf D#1} (19#2) #3}
\def\prep#1,#2,#3{         {\it Phys. Rep.\/ }{\bf #1} (19#2) #3}
\def\prl#1,#2,#3{          {\it Phys. Rev. Lett.\/ }{\bf #1} (19#2) #3}
\def\pro#1,#2,#3{          {\it Prog. Theor. Phys.\/ }{\bf #1} (19#2) #3}
\def\rmp#1,#2,#3{          {\it Rev. Mod. Phys.\/ }{\bf #1} (19#2) #3}
\def\sjnp#1,#2,#3{         {\it Sov. J. Nucl. Phys.\/ }{\bf #1} (19#2) #3}
\def\sp#1,#2,#3{           {\it Sov. Phys.-Usp.\/ }{\bf #1} (19#2) #3}
\def\zp#1,#2,#3{           {\it Zeit. f\"ur Physik\/ }{\bf #1} (19#2) #3}
\def\etal{ {\it et al.}}
\def\ie{{\it i.e.} }

\begin{document}
\thispagestyle{empty}
\setcounter{page}{0}
\renewcommand{\thefootnote}{\fnsymbol{footnote}}

\begin{flushright}
CERN-TH/95--250 \\[2ex]
{\large \bf hep-ph/9511466} \\
\end{flushright}

\vspace{5ex}
\begin{center}

{\Large \bf $R$-Parity Violation at LEP2 : Virtual Effects}\\

\bigskip
\bigskip
{\sc
   Debajyoti Choudhury\footnote{debchou@surya11.cern.ch}
   }

\bigskip
{\it Theory Division, CERN, CH--1211 Gen\`eve 23, Switzerland.}\\
\bigskip
\bigskip
{\bf Abstract}
\end{center}

\begin{quotation}
We investigate the ability of LEP2 to detect possible $R$-parity
violation, especially for the case where direct production
cross-sections are too small for all superparticles. We demonstrate
that for coupling strengths allowed by present experiments,
sfermion-exchange diagrams can contribute significantly to the
$e^+ e^- \rightarrow f \bar{f} $ process. This would be a useful
tool in further constraining the parameter space. Similar arguments
hold for leptoquarks and dileptons as well.
\end{quotation}

\vspace{170pt}
\noindent
\begin{flushleft}
CERN-TH/95--250\\
October 1995\\
\end{flushleft}

\newpage
\setcounter{footnote}{0}
\renewcommand{\thefootnote}{\arabic{footnote}}

Supersymmetry is perhaps the most popular and promising theoretical
concept going beyond the Standard Model (SM). The
 Minimal Supersymmetric Standard Model (MSSM) \cite{mssm}
is obtained from the SM by the naive supersymmetrization of both the
particle content and the couplings in the latter. However, the most
general Lagrangian consistent with supersymmetry as well as with the
$SU(3)_c \otimes SU(2)_L \otimes U(1)_Y$ gauge symmetry contains
terms that have no analogue within the SM. This comes about as one of the
Higgs supermultiplets has the same quantum numbers as the doublet lepton
superfields and thus may be replaced by the latter
 in the Lagrangian. A similar
effect may also be spontaneously generated if any of the sneutrino
fields develops a vacuum expectation value. Furthermore, trilinear terms
involving the singlet quark superfields are also allowed.
The additional pieces in the superpotential may thus
be parametrized as \cite{rpar}
\be
    {\cal W}_{\not R} =  \lambda_{ijk} L_i L_j E^c_k
                        +  \lambda'_{ijk} L_i Q_j D^c_k
                        +  \lambda''_{ijk} U^c_i D^c_j D^c_k \ ,
      \label{R-parity}
\ee
where $E^c_i, U^c_i, D^c_i$
are the singlet  superfields and
$L_i$ and $Q_i $ are the $SU(2)$-doublet lepton and quark
superfields.  The coefficients $\lambda''_{ijk}$
are antisymmetric under the interchange of the last two indices,
while $\lambda_{ijk}$ are antisymmetric
under the interchange of the first two.

Terms as in eq.(\ref{R-parity}) obviously have striking phenomenological
consequences. For example, the $\lambda_{ijk}$ and $\lambda'_{ijk}$
couplings violate lepton number, while the $\lambda''_{ijk}$ violate
baryon number. Simultaneous presence of both
may therefore lead to catastrophically
high rates for proton decay. In a similar vein, pairs of such
couplings may induce tree-level flavour-changing neutral currents.
Such considerations led to the  introduction of a discrete symmetry
known as ``$R$-parity''. Expressible in terms
of baryon number ($B$), lepton
number ($L$) and the intrinsic spin ($S$),
$R \equiv (-1)^{3B + L + 2 S}$ has a value of $+1$ for
all SM particles and $-1$ for all their superpartners.
Apart from ruling out each of the terms in eq.(\ref{R-parity}),
an exact $R$-parity has the additional
consequence of rendering the lightest superpartner (LSP)  stable.

While this discrete symmetry may be phenomenologically desirable,
there is no clear theoretical motivation for it to exist. Even
phenomenologically, an exact $R$-parity is an overkill. The constraints
from proton-decay may be evaded by assuming that all of the
$\lambda''_{ijk}$ vanish identically\footnote{In fact, only some of these
need to be vanishingly small. The constraints on the others are somewhat
weaker \protect\cite{probir}.}. Such an assumption is well motivated within
certain theoretical scenarios \cite{hall-suzuki} and we shall hold
it to be true. The problem of preservation of GUT-scale
baryogenesis \cite{baryo} is thus rendered largely ineffectual.
The presence of the other \rp\  terms can, however,
also affect the baryon asymmetry of the universe.
It has been argued \cite{dreiner}, though,
that such bounds are highly model-dependent and  can
hence be evaded, for example by conserving one lepton flavour
over  cosmological time scales.

Couplings as in eq.(\ref{R-parity}) can also be bounded by using various
low-energy data such as lepton- or meson-decays \cite{barger,d_tau}, or
from analyses of the $Z$-decay modes at LEP \cite{z-decay}. Many of these
constraints are relatively weak though, and the allowed magnitudes for
the corresponding couplings may lead to remarkable signals at
LEP2 \cite{grt,bkp,lep2}. Most of such studies concentrate on scenarios wherein
the LSP and/or other supersymmetric particles are light enough to be
produced at LEP2. The breaking of $R$-parity then leads to certain
tell-tale signatures.

In this Letter, we investigate the orthogonal
set, namely we assume that {\em none} of the supersymmetric particles
(including the LSP) can be produced with a significant cross-section.
The suppression could come from two sources : ($i$)  all
supersymmetric particles are relatively
heavy, or ($ii$) the  lighter ones couple very weakly to the
relevant SM particles.
Any possible effect can then only be virtual. The best testing ground at
LEP2 is provided by pair-production of light charged fermions :
\be
    e^+ e^- \rightarrow f \bar{f} \ .
    \label{pair-prodn}
\ee
For experimental reasons, we confine ourselves to $f = e, \mu, \tau, b , c$,
and
hence to the couplings $\lambda_{12k, 13k, 231}$ and $\lambda'_{113, 12k}$.
The SM contribution to the above process is in the form of
$\gamma,Z$ mediated $s$-channel diagrams (for $f = e$, additional
 $t$-channel ones too).
The introduction of the  terms in eq.(\ref{R-parity}) leads to new
$t$-channel (also $s$-channel for $f = e$)
diagrams governed by the following Yukawa couplings :
\be
\barr{rcl}
    {\cal L}_{\lambda,\lambda'} & = & \displaystyle
    \lambda_{ijk}  \left[{\tilde\nu}_{iL} \bar
e_{kR} e_{jL} + {\tilde e}_{jL} \overline{e_{kR}} \nu_{iL} + {\tilde
e}^\ast_{kR} \overline{(\nu_{iL})^C} e_{jL} - (i \leftrightarrow
j)\right] + {\rm h.c}
\\[2ex]
& + & \displaystyle
\lambda'_{ijk} \left[ \hspace*{.5em}
                       {\tilde \nu}_{iL} \overline{d_{kR}} d_{jL}
                     + {\tilde d}_{jL} \overline{d_{kR}} \nu_{iL}
                     + {\tilde d}^\ast_{kR} \overline{(\nu_{iL})^C} d_{jL}
               \right.
\\[2ex]
&  & \displaystyle \left. \hspace*{1.55em}
                     - \: {\tilde e}_{iL} \overline{d_{kR}} u_{jL}
                     - {\tilde u}_{jL} \overline{d_{kR}} e_{iL}
                     - {\tilde d}^\ast_{kR} \overline{(e_{iL})^C} u_{jL}
         \right] + {\rm h.c.}
\earr
\ee
For $f = e, \mu, \tau$  we then have sneutrino-mediated diagrams, while
for $f = b,c$ we have ${\tilde u}_{jL}$- and ${\tilde d}_{jR}$
exchanges respectively. At this stage it is useful to note that
if we close our eyes to the global quantum numbers for the scalars,
the squark interaction mimics that of certain scalar leptoquarks while
the sneutrinos mimic a dilepton. Thus much of the analysis presented
here can trivially be extended to a discussion of leptoquark and
dilepton couplings.

Since the \rp interactions have a  structure different from
the SM one, the  angular distribution for the $f \bar{f} $ pair
is a sensitive probe for the existence of the former.
In fact, this very difference alters the mass-dependence of
these bounds from the linear relation derived from low-energy
measurements.
The sensitivity of the experiment can be gauged by
dividing  the angular width of the experiment into
bins and comparing the observed number of events $n_j$ in each with the SM
prediction $n_j^{SM}$. A  quantitative measure is given by a
$\chi^2$ test\footnote{Since the expected number of events is
relatively large, we do not envisage a qualitative improvement by
adopting an unbinned maximum-likelihood test. However, this should
be checked with a full detector simulation.}:
\be \displaystyle
    \chi^2 = \sum_{j=1}^{ {\rm bins} }
      \left( \frac{ n_j^{SM} - n_j}{\Delta n_j^{SM} } \right)^2\ .
\label{chi2}
\ee
The number of events is obtained by integrating the differential
distribution over the angular bin and is given by
\be
    n_i = \sigma_i \epsilon L\ .
\ee
where $\epsilon$ is the detector efficiency and $L$ is the machine luminosity.
The error in eq.(\ref{chi2}) is obtained by adding the
 statistical and systematic ones in quadrature :
\be
\Delta n_j^{SM} = \sqrt{ (\sqrt{n_j})^2 + (\delta_{\rm syst} n_j)^2 }.
  \label{error}
\ee

To be quantitative, we shall use
\be
\sqrt{s} = 192 \GeV \qquad L = 500 \ {\rm pb}^{-1} \qquad
\delta_{\rm syst} = 0.02 \ .
       \label{machine}
\ee
For experimental convenience, we demand that the outgoing leptons or jets
in eq.(\ref{pair-prodn}) be at least $20^\circ$ away from the beam pipe.
Within this restricted region we assume uniform detection
efficiencies~\cite{tariq}
\be
\epsilon_e = \epsilon_\mu = 0.95 \qquad \epsilon_\tau = 0.90
\qquad
\epsilon_b = 0.25 \qquad \epsilon_c = 0.05.
\ee

Dividing the above-mentioned angular region (between $20^\circ$ and
$160^\circ$) into 20 equal-sized
bins\footnote{We find that the sensitivity of the results to the binning
is rather weak for bin cardinality between 15 and 30.},
we then perform the $\chi^2$ test as in
eq.(\ref{chi2}). To avoid spurious contributions to the $\chi^2$
we reject a bin from the  analysis if either
$(i)$    the difference between the SM expectation
        and the measured number of events is less than 1
or
$(ii)$   the SM expectation is less than 1 event
        while the measured number is less than 3.

We present our results in the form of 95\% C.L. bounds in the two-parameter
space of the sfermion mass and  the \rp\ coupling.
The interpretation is straightforward. Any combination of the two
parameters {\em above} the curves (\ie away from the origin) can be ruled
out at 95\% C.L.\footnote{If the value of one of the parameters were known,
then a 98.6\% C.L. bound on the other would be
 given by the projection on the corresponding  axis.}.
As we are dealing with one-sided bounds,
this corresponds to $\chi^2 > 4.61$ in eq.(\ref{chi2})~\cite{pdg}.

\begin{figure}[h]
\vskip 4.2in\relax\noindent\hskip -1.8in\relax{\includegraphics{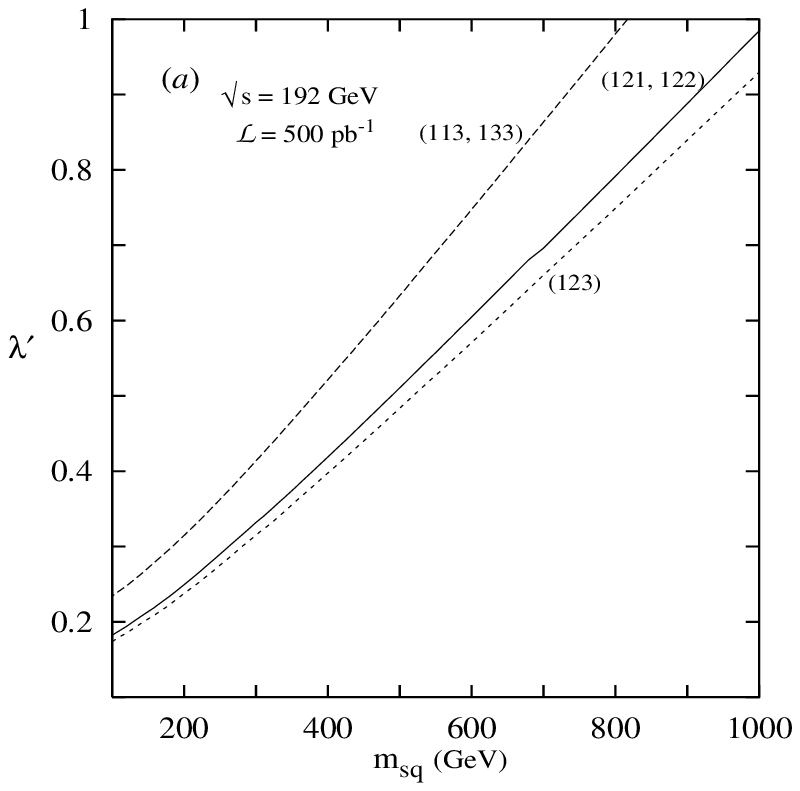}}
          \relax\noindent\hskip 3.3in\relax{\includegraphics{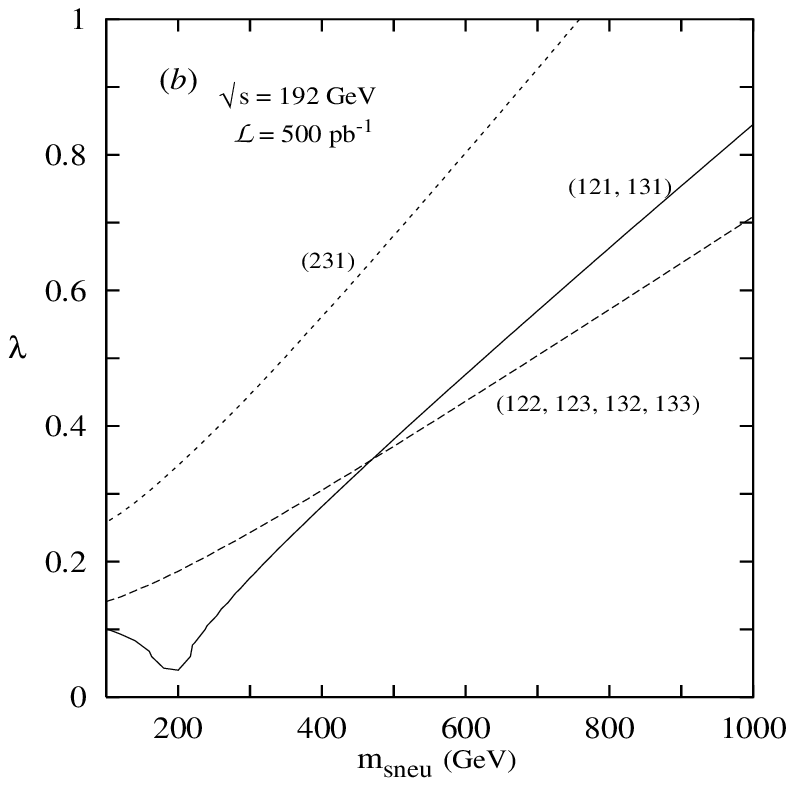}}
\vspace{-18ex}
\caption{ {\em  Contours of detectability in the coupling--sfermion mass plane.
The
numbers in the parentheses refer to the indices on $\lambda'$ ($\lambda$).
The parameter space {\em above} the curves can be ruled out at 95\% C.L. For
fixed value of one parameter, the projection onto the other axis defines
a 98.6\% C.L.} }
\label{fig:constraint}
\end{figure}
Figure (1$a$) shows the bounds for all the relevant $\lambda'$s as
a function of the squark-mass\footnote{Where more than one sfermion can
contribute,
we assume them to have identical mass.}.
Since both $\lambda'_{121}$ and $\lambda'_{121}$ are probed by the
angular distribution of charm jets, the sensitivity is identical.
The constraints on $\lambda'_{123}$ is somewhat stronger as both the
$c$ and the $b$ quarks are observable in the final state.
A curious point is that inspite of lower detector efficiency,
$c$-jet distributions are more sensitive to
the presence of \rp\ couplings than the $b$-jets. This can be traced
to the relative size of the interference term.
The bounds on $\lambda'_{121}$ and $\lambda'_{123}$ are
significantly stronger than those available today \cite{barger,d_tau}.
Though the direct experimental bounds on $\lambda'_{122}$ and
$\lambda'_{133}$ as weak, these induce radiative correction
to the Majorana mass for the electron
neutrino~\cite{mohap,grt,d_tau} and thus can be restricted severely.
On the other hand, $\lambda'_{113}$  is tightly bound from
charged current universality \cite{barger}.

Figure (1$b$) shows similar bounds for the $\lambda$s as a function
of the sneutrino mass. While
$\lambda_{122}$ or $\lambda_{132}$ lead to a $t$-channel sneutrino
 ($\tilde\nu_{\mu L}$ and $\tilde\tau_{\mu L}$ respectively) diagram
for $e^+ e^- \to \mu^+ \mu^-$, $\lambda_{123}$ or $\lambda_{133}$ do the
same for $\tau$-production. Since the detector
efficiencies for $\mu$-pair and $\tau$-pair are quite similar, the
corresponding bounds are almost indistuingishable from each other.
Though $\lambda_{231}$ leads to additional diagrams in both the $\mu$--
and $\tau$-channels, the sensitivity. This again can be traced back to
the size of the interference term (since this coupling sees $e_R$ rather
than $e_L$). The most interesting cases are however those of
$\lambda_{121}$ and $\lambda_{131}$. Apart from muonic (tauonic) final
states, these couplings contribute to $e^+ e^- \to e^+ e^- $ as well.
The dip in the contour corresponds to a resonance production of a
sneutrino which subsequently decays into a $e^+ e^- $
pair\footnote{Provided a neutralino or a chargino is lighter than this
sneutrino,
part of this particular parameter range may also be investigated by
looking for decays into these channels~\protect\cite{lep2}.}.
As in the $\lambda'$ case, here too non-observation of a Majorana mass
for the $\nu_e$ constrains $\lambda_{122, 133}$. Of the rest,
these new bounds on $\lambda_{131, 132}$ would be the strongest yet.
The others are at best similar to the present ones. Only in the case
of the sneutrinos being considerably lighter than the corresponding
charged sleptons, will there be a qualitative improvement.

To conclude, we point out that the present constraints on some of the
\rp\ couplings are relatively weak. If the supersymmetric particles are
light enough to be produced copiously at LEP2, we would shortly
be in a position to witness dramatic signals. On the other hand, if they are
either
too heavy or too weakly coupled to be produced, indirect effects provide
us with a means to investigate this sector. We exhibit that quite a few
of the lepton-number violating \rp\
 couplings lead to significant deviations in the
$e^+ e^- \rightarrow f \bar{f} $ angular distributions. For some
of the couplings, this effect can
be used to impose bounds that are stronger than any available today,
while for others it will provide a complementary test.
A similar analysis would also be applicable to a large class
of leptoquark and dilepton couplings.

 {\bf Acknowledgements}: I wish to thank T.~Aziz and G. Bhattacharyya
for  useful discussions.

\end{document}